\documentclass{elsart}
\usepackage[dvips]{graphicx}
\usepackage{dcolumn}
\usepackage{epsfig}

\begin{document}

\begin{frontmatter}

\title{Neutrino-nucleus reactions in the delta resonance region}

\author[label1,label3]{B. Szczerbinska}
\author[label2]{T. Sato}
\author[label3]{K. Kubodera}
\author[label4]{T. -S. H. Lee}
\address[label1]
{Dakota State University, Madison, South Dakota 57020, USA}
\address[label2]
{Department of Physics, Osaka University, Toyonaka, Osaka 560-0043,
 Japan}
\address[label3]
{Department of Physics and Astronomy, University of South Carolina,
Columbia, South Carolina 29208, USA}
\address[label4]
{Physics Division, Argonne National Laboratory, Argonne, Illinois 60439, USA}

\begin{abstract}

Reliable estimates of neutrino-nucleus reactions in the 
resonance-excitation region play an important role in many of 
the on-going and 
planned neutrino oscillation experiments.
We study here neutrino-nucleus reactions in the delta-particle
excitation region with the use of neutrino pion-production amplitudes
calculated in a formalism in which the resonance 
contributions and the background amplitudes are treated on the same footing.
Our approach leads to the neutrino-nucleus reaction cross sections 
that are significantly different from those obtained in the conventional
approach wherein only the pure resonance amplitudes are taken into account.
To assess the reliability of our formalism, we calculate the electron-nucleus
scattering cross sections in the same theoretical framework; the calculated 
cross sections agree reasonably well with the existing data.

\end{abstract}

\begin{keyword}
\PACS 13.15.+g \sep 13.60.Le \sep 25.30.Rw \sep 25.30.Pt
\end{keyword}
                                                                   
\end{frontmatter}

\section{Introduction}

It is well recognized that the precise knowledge of neutrino-nucleus 
reaction cross sections is of importance in analyzing neutrino oscillation 
experiments; for recent reports, see {\it e.g.} 
Refs.~\cite{sakuda,Tanetal06,Wasetal06,Pet06,Beletal06}.
In particular, neutrino-nucleus reactions at incident 
neutrino energies around 1~GeV play a prominent role
in many cases including the experiments 
at K2K~\cite{Tanetal06}.
To obtain estimates of the relevant cross sections,
one must at present rely on theory,
and much theoretical effort has been invested 
to provide these 
estimates~\cite{Koletal03,Benetal05,Nieetal05,gil97,Satetal06,Leietal06}.
In an attempt to make a quantitative estimation of 
neutrino-nucleus reaction cross sections,
it is useful to study simultaneously
the related electron-nucleus reactions
within the same general theoretical framework,
and this strategy has been pursued by many authors.
In electron-nucleus scattering in the GeV region,
quasi-elastic scattering and pion-production processes are 
known to be the main reaction mechanisms,
and similar features are expected to manifest themselves
also in the neutrino-nucleus reactions in the GeV region.

For quasi-elastic scattering, 
the relevant transition operators are essentially known, so
the main theoretical issue is how to incorporate
various nuclear effects for the initial and final states.
The early works were based on the Fermi gas model~\cite{moniz,kuretal90},
but recent investigations incorporate the nuclear correlation effects in 
the initial state with the use of the spectral function
and take account of the final-state interactions 
on the outgoing nucleon~\cite{Benetal05}.
As regards the pion-production process, 
in addition to these nuclear effects, 
the structure of the transition operators responsible for
pion production needs to be carefully studied.
These operators can in principle involve more than one nucleon,
but it is in general expected that pion production on a single nucleon
should give a dominant contribution.
Neutrino-induced pion production on the nucleon
in the resonance region has been studied so far mostly 
with the use of pure resonance excitation amplitudes.
In some studies these amplitudes were evaluated
in the quark model, see, e.g., Ref.~\cite{RS81}.
In recent studies by Paschos and his 
collaborators~\cite{pascos1,lp06},
the resonance excitation amplitudes due to the vector current
were directly related to the empirically known electro-excitation
amplitudes, while those due to the axial-vector current
were constrained by invoking PCAC.

Meanwhile, it is to be noted that pion production can take place
not only through resonance excitations but also via non-resonant processes.
Two of the present authors~\cite{sl1,sl2}
have recently developed a dynamical model 
for describing photo- and electro-production of pions off the 
nucleon around the $\Delta$-resonance region,
with the view to systematically incorporating  
both the resonance and non-resonance contributions. 
Hereafter we refer to this approach 
as the SL-model (the Sato-Lee model).
The development of the SL-model was motivated by 
recent extensive experimental studies
of electron- and photon-induced meson-production reactions
on the nucleon in the resonance region.
The main objective of these experiments 
is to study the non-perturbative features of QCD 
by testing the resonance properties 
as predicted by QCD-inspired models and/or lattice simulations.
The SL-model was subsequently extended to
weak-interaction processes~\cite{sl3,sl4}, 
and it was shown that this model gives a successful description 
of neutrino-induced pion production in the $\Delta$-resonance region. 

As explained in more detail later, 
the SL-model starts from the non-resonant meson-baryon
interaction and the resonance interaction, and the 
unitary amplitudes are obtained from the scattering equation.
It leads to fairly consistent descriptions of all the available data 
for the electroweak reactions in the $\Delta$-resonance region.
It has been shown that treating the resonance and non-resonance 
amplitudes on the same footing can have significant observable
consequences.
In particular, the inclusion of  the pion cloud effects as considered in SL
can resolve a long-standing puzzle that 
the $N$-$\Delta$ magnetic dipole transition form factor $G_M$ 
predicted by the quark model is
smaller than the empirical value by as much as  $\sim$40\%.
Furthermore, the electric E2($G_E$) and
Coulomb C2($G_C$) form factors for the $N$-$\Delta$ transition
in electron scattering 
calculated in the SL-model show pronounced 
momentum dependences
due to the pion cloud effects, 
which suggests non-negligible deformation effects 
in the $N$-$\Delta$ transition.
Regarding the neutrino reactions, a serious problem 
that has been known for quite some time is that the axial-vector 
$N$-$\Delta$ transition strength calculated 
in the constituent quark model~\cite{Hemetal95}
is lower than the empirical value~\cite{Kitetal90}
by about 35 \%.
It is noteworthy that the dynamical pion cloud effects
included in the SL-model~\cite{sl3}
can naturally remove this discrepancy.

In view of these successes,
it seems worthwhile to study neutrino-nucleus reactions in the
resonance region with the use of the SL-model amplitudes
for neutrino-induced pion production on the nucleon.
We describe here our first attempt at such a study
and present the cross sections,
the energy spectrum of the final lepton
(for charged-current reactions), and
the lepton-momentum transfer distribution.
Our work is basically of exploratory nature
and, as far as the nuclear effects are concerned,
we only consider those that can be taken into account 
with the use of a modified Fermi gas model 
wherein nuclear correlations are approximately subsumed  
into the spectral function~\cite{Benetal05}.
Despite these limitations, our investigation
is hoped to be informative as the first calculation
of neutrino-nucleus reactions in the $\Delta$-resonance region
based on the electroweak pion-production amplitudes 
calculated in SL~\cite{sl1,sl2}, whose validity has been extensively 
tested by the Jlab data~\cite{fro,joo}.
It is understood that, as the experimental precision improves,
more detailed calculations will be called for
that incorporate higher order effects.
In particular, the final-state interaction (FSI) must be treated properly. 
As discussed in Ref.~\cite{Benetal05} and many
earlier works on inclusive electron scattering,
FSI re-distributes the
inclusive cross sections and,
for the incident electron energy around 1 GeV,
FSI can reduce the strength 
at the quasi-free peak by about 10 $\%$.
We remark that
a detailed study~\cite{Horikawa} indicates that
the FSI effects for inclusive reactions, properly treated,
can be rather different from those for
exclusive processes~\cite{martinez06}.
Meanwhile, in the pion production region, 
we need to take into account 
pion absorption and medium effects on $\Delta$ propagation.
To this end, one may profitably use the information obtained 
in the well-developed $\Delta$-hole model~\cite{Lenz-Thies};
such a study has been made in Ref.~\cite{cassing06} 
within the framework of the dynamical transport approach. 
Our present calculation, however, falls short of considering FSI. 
Since the importance of these FSI effects grows 
rather fast with the increasing target mass number 
(this is particularly true for pion absorption),
we limit ourselves here to nuclear targets of 
low mass numbers 
and concentrate on the A=12 target.
Although heavier nuclei such as $^{56}Fe$
are important in some neutrino-oscillation experiments~\cite{Beletal06},
we can deal with these cases only after  
FSI is incorporated into our formalism.

\section{Sato-Lee (SL) Model}

As the SL-model has been fully described 
in Refs.~\cite{sl1,sl2,sl3,sl4},
we give here only a brief explanation of the model,
using as an example the case of pion photoproduction.
The effective Hamiltonian $H_{eff}$ in the SL-model for
this process is given by
\begin{eqnarray}
H_{eff}  =  H_0  + v_{\pi N} + v_{\gamma \pi} +
 \Gamma_{\pi N \leftrightarrow \Delta} +
 \Gamma_{\gamma N \leftrightarrow \Delta}\,,
\end{eqnarray}
where $H_0$ is the free Hamiltonian;
 $v_{\pi N}$ and $v_{\gamma \pi}$  represent
the non-resonant pion-nucleon and pion photoproduction
interactions, respectively, while
$\Gamma_{\pi N \leftrightarrow \Delta} $
and $\Gamma_{\gamma N \leftrightarrow \Delta}$
are responsible for the creation and annihilation
of a bare $\Delta$-resonance.
By solving the Lippmann-Schwinger equation
based on the above effective Hamiltonian,
we obtain the amplitude for pion production on a nucleon as
\begin{eqnarray}
  T_{\gamma \pi}& = &
t_{\gamma \pi}(E) + 
\frac{\bar{\Gamma}_{\Delta \rightarrow \pi N}(E){\bar{\Gamma}_{\gamma N
\rightarrow \Delta}}(E)}{E - m_\Delta^0 - \Sigma(E)}, \label{tmat}
\end{eqnarray}
where $E$ is the total energy of  the pion and nucleon 
in the center-of-mass system.
The first term $t_{\gamma\pi}$ is the non-resonant amplitude,
which arises from the vertices $v_{\pi N}$ and $v_{\gamma N}$ alone,
while the second term represents the resonant amplitude
involving the dressed vertex $\bar{\Gamma}$.
We note that the bare resonance vertex $\Gamma$ is renormalized into
$\bar{\Gamma}$  by the non-resonant meson cloud effects arising
from rescattering as
\begin{eqnarray}
\bar{\Gamma}_{\gamma N\rightarrow \Delta}(E)
& = & \Gamma_{\gamma N\rightarrow \Delta}
  +  \Gamma_{\pi N\rightarrow \Delta}G_0 t_{\gamma \pi}(E),
\end{eqnarray}
and that the renormalized resonance vertex
$\bar{\Gamma}$ should exhibit a significant deviation
from
the bare vertex $\Gamma$
because of the meson cloud effects.
In conventional analyses, however,
one disregards the difference between the bare and 
renormalized resonance vertices
and, assuming the Breit-Wigner form
for (what in our approach is identified as) 
the renormalized vertex,
tries to extract the parameters characterizing that form
by fitting to the data. 
However, as discussed in detail in Refs.~\cite{sl1,sl2,sl3,sl4}
and as briefly mentioned in the introduction,
the resonance properties deduced from this simplified treatment
tend to exhibit significant discrepancies with the theoretical predictions,
indicating the importance of considering the resonant and non-resonant
contributions simultaneously.

\begin{figure}[htb]
\includegraphics[width=4.5cm]{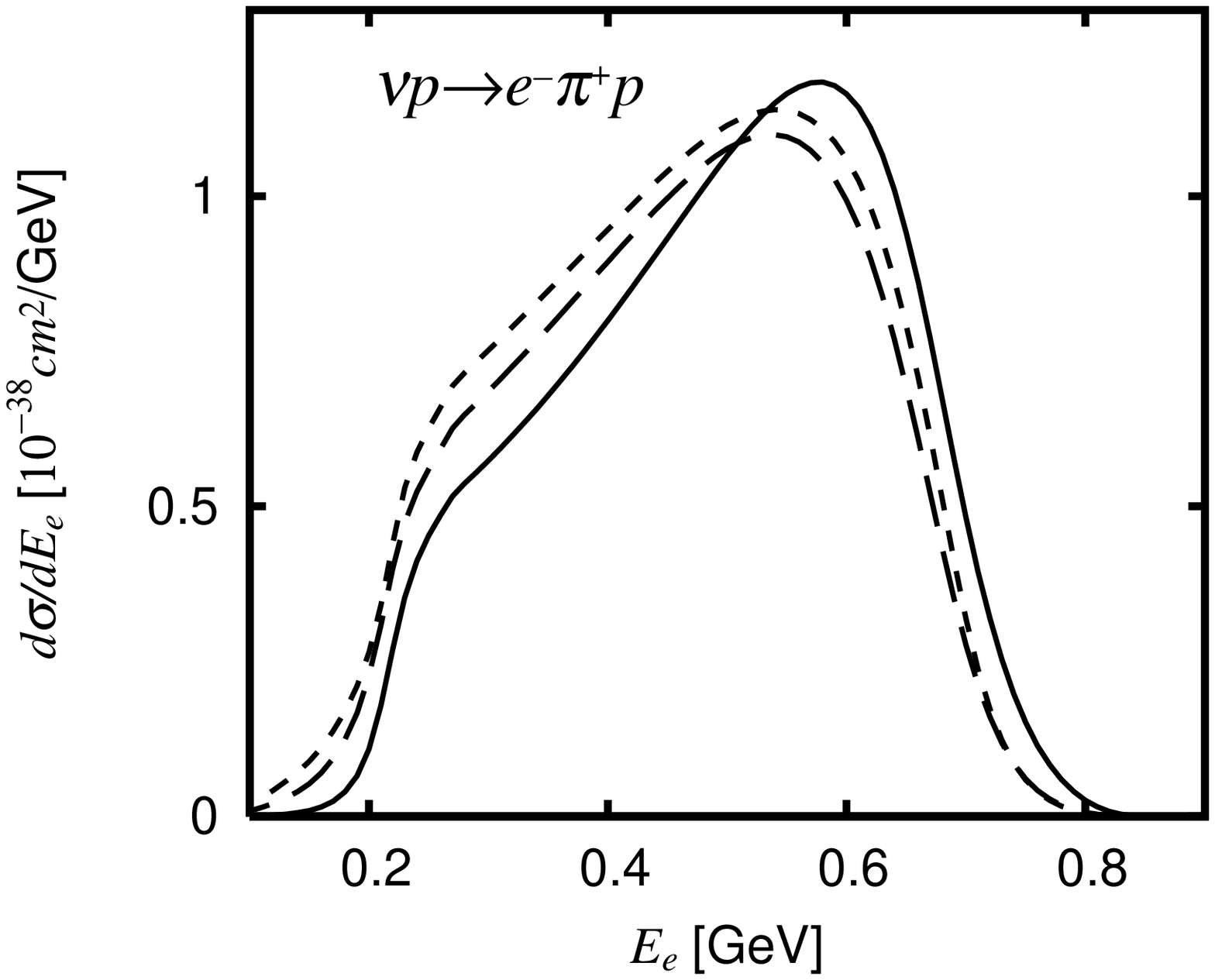}
\includegraphics[width=4.5cm]{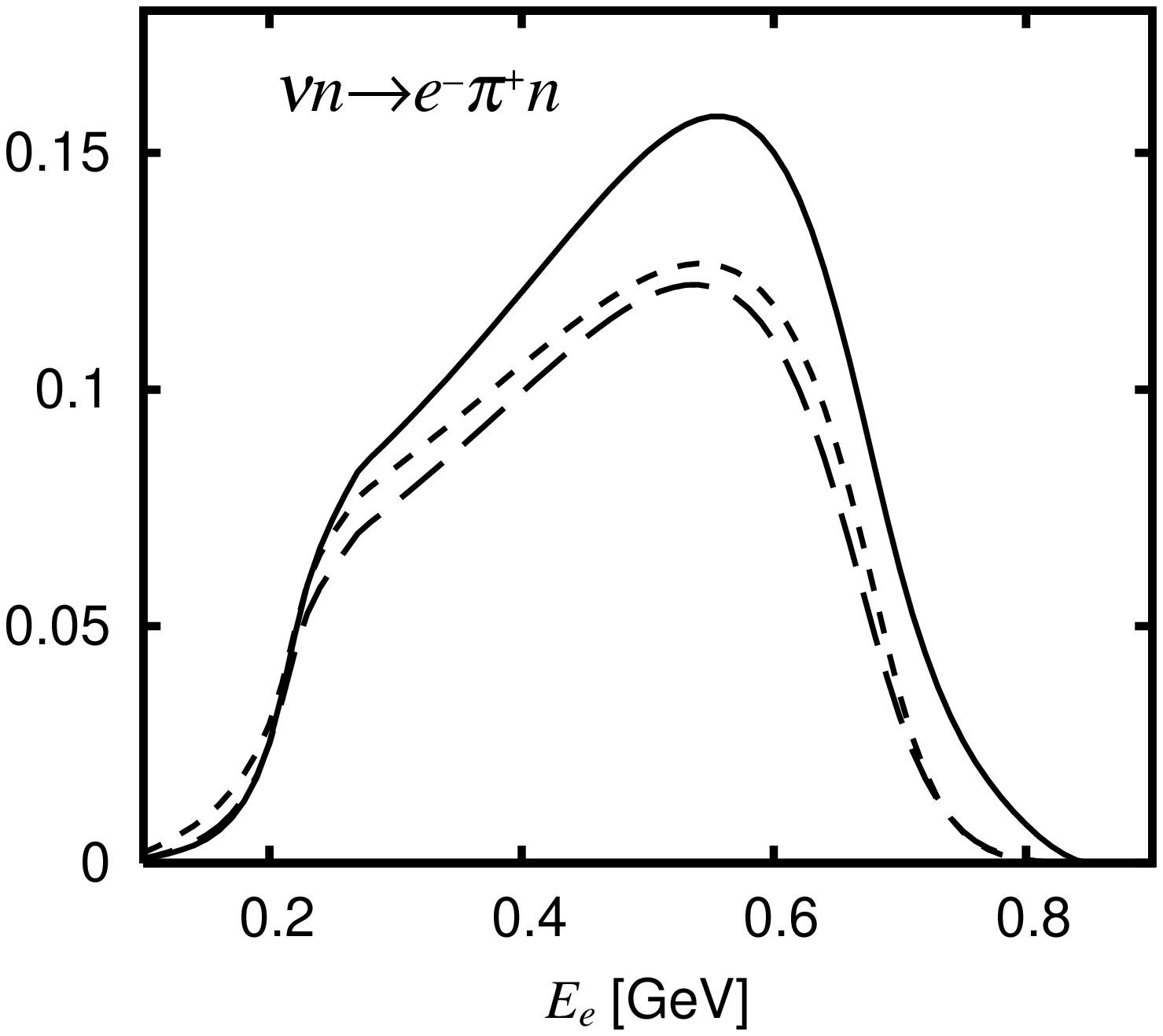}
\includegraphics[width=4.5cm]{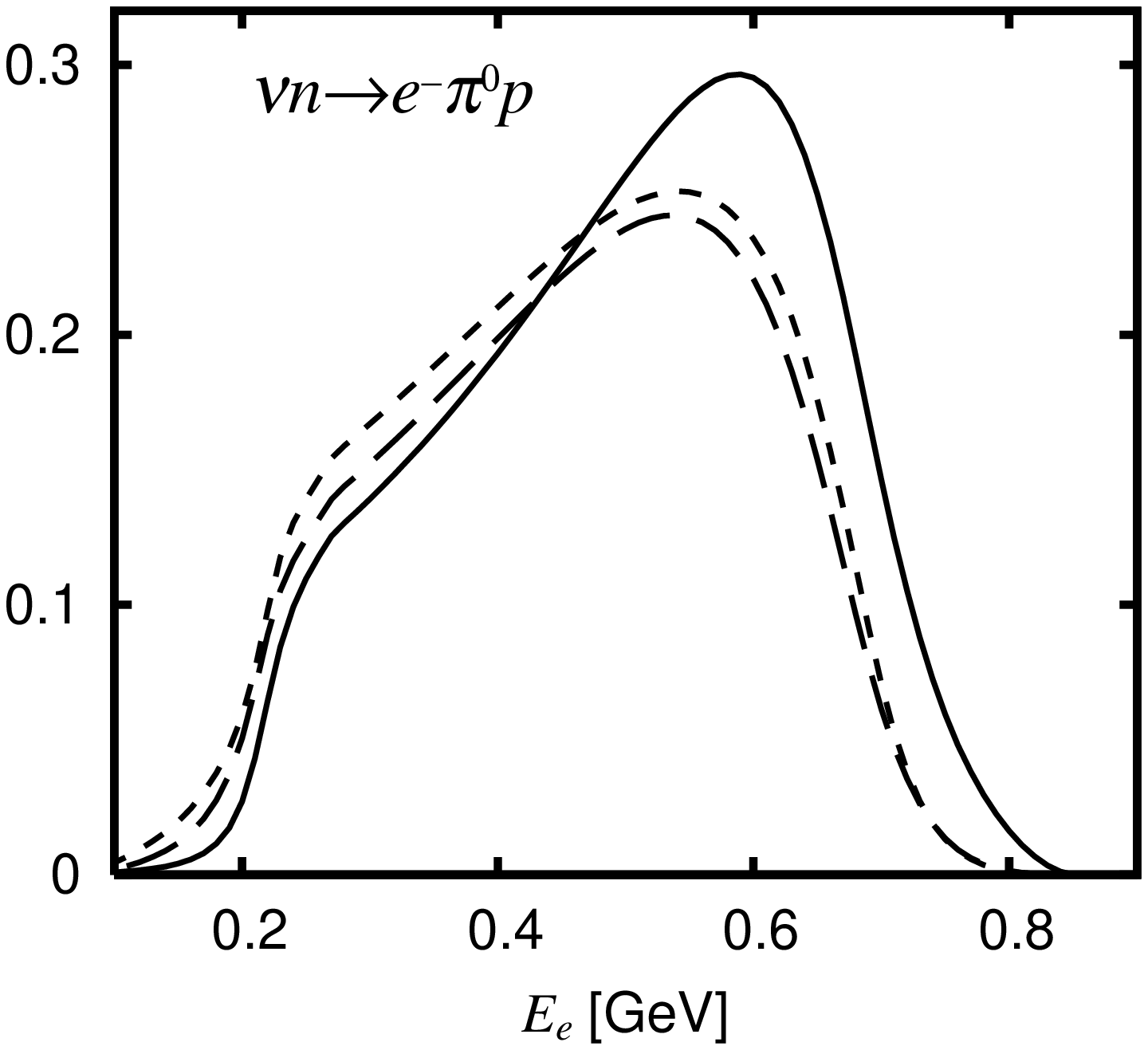}
\caption{\label{fig-h} Electron energy spectrum for
 $ \nu_e p \rightarrow e^-  \pi^+ p$ (left panel),
 $ \nu_e n \rightarrow e^-  \pi^+ n$ (middle panel), and
 $ \nu_e n \rightarrow e^-  \pi^0 p$ (right panel).
 For the explanation of the three curves in each panel, see the text.}
\end{figure}

We expect that the unified treatment of 
the resonant and non-resonant contributions
should have significant consequences in the neutrino-nucleon and
neutrino-nucleus reactions as well.
To illustrate this point, we give in Fig.~\ref{fig-h}
the electron energy spectrum $d\sigma/dE_e$ for  the
$\nu_e  N \rightarrow e^- \pi N$ reaction
calculated in the SL-model.
Since the isospin of the $\pi N$ system 
for the $\nu_e p \rightarrow e^- \pi^+ p$  reaction is $3/2$,
one may naively expect that the  $\Delta$-resonance amplitude
dominates the cross section. 
However, the energy spectra obtained in our SL-model calculation
(shown in the solid lines) are markedly different from 
those obtained from
the dressed resonant amplitudes (shown in the long-dash lines). 
For comparison,  the results obtained 
in the Lalakulich-Paschos (LP) model~\cite{pascos1}
are also shown in  the short dashed lines.
If in our approach we drop the contribution of the 
non-resonant amplitudes (retaining only the resonance contributions),
the results turn out to be very similar
to those obtained in the LP-model.

\begin{figure}
\includegraphics[width=6cm]{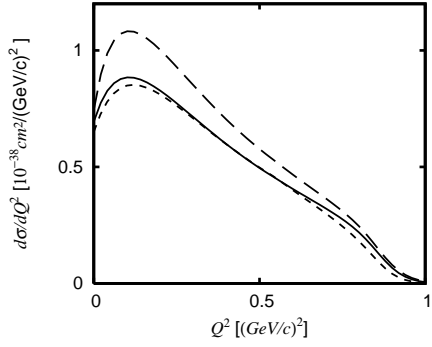}
\caption{\label{fig-dq-n} Differential cross sections $d\sigma/dQ^2$ for
$ \nu_e N \rightarrow e^-\pi N$ at $E_\nu=$ 1~GeV,
where $Q^2= -q^2$ and $q\equiv p_e-p_\nu$ is the lepton momentum transfer.
Solid line -- proton target (SL-model calculation); 
short-dashed line -- proton target  (resonance amplitude only);
long-dashed line -- neutron target case multiplied with a factor of 3
(SL-model calculation).
The final $\pi N$ state is $\pi^+p$ for the proton target, 
whereas both $\pi^0 p$ and $\pi^+n$ can contribute in the neutron target case.}
\end{figure}

Since neutrino-nucleus reactions obviously involve 
both $\nu p$ and $\nu n$ reactions,
it is also informative to examine how the non-resonant 
contributions can affect the relative importance of 
the $\nu p$ and $\nu n$ contributions.
If the $I=3/2$ resonance amplitude dominates, 
the cross section on the neutron should be 
$1/3$ of that for the proton.  
Fig.~\ref{fig-dq-n} shows $d\sigma/dQ^2$ for the proton target
(solid line) and $3\times d\sigma/dQ^2$ for the neutron target 
(long-dashed line) for $E_\nu=$ 1 GeV. 
The curves should agree with each other
if the delta mechanism dominates. 
However, the neutron cross section
is about 20\% larger than the value expected from $\Delta$ dominance.
The short-dashed line gives $d\sigma/dQ^2$ for the proton target
obtained with the use of the dressed resonance amplitude alone.
We note that this curve overlaps rather well with
the solid line corresponding to the SL-model results,
a feature to be contrasted with the behavior of 
$d\sigma/dE_l$ shown in Fig.~\ref{fig-h}.
We learn from this that, even for the same reaction,
some observables are more sensitive than others 
to the difference between the SL-model results and
those of the resonance-contribution-only approach.

\section{Neutrino-nucleus reaction}

We consider the charged-current (CC) neutrino-nucleus reaction
\begin{eqnarray}
\nu_{\ell}(p_\nu)+|i(P_i)\!>
\rightarrow \ell(p_\ell)+|f(P_f)\!>\,,
\end{eqnarray}
where $|i\!>$ represents a target nucleus of mass $A$,
$|f\!>$ stands for a final hadronic state, and
$\ell$ is a lepton flavor ($\ell=e,\,\mu\,,\tau$);
the relevant momenta are indicated in the parentheses.
The cross section for this process  
is given in terms of the lepton tensor $L_{\mu\nu}$  and 
the hadron tensor  $W^{\mu\nu}$ as
\begin{eqnarray}
\frac{d\sigma}{dE_l d\Omega_l} & = & \frac{p_l}{p_\nu}
\frac{G_F^2 \cos \theta_c^2}{8\pi^2}L_{\mu\nu}W^{\mu\nu}.
\end{eqnarray}
The lepton tensor is expressed as
\begin{eqnarray}
L^{\mu\nu} = 2\,[\,p_{l}^\mu p_\nu^\nu + p_{l}^\nu p_\nu^\mu - g^{\mu\nu}
 (p_\nu\cdot p_{l} -
 m_l^2) + i \epsilon^{\mu\nu\alpha\beta} p_{\nu, \alpha} p_{l,\beta}\,]\,,
\end{eqnarray}
where $m_\ell$ is the mass of the final lepton.
The hadron tensor is related to the matrix elements
of the hadronic weak current $J^\mu$ as
\begin{eqnarray*}
W^{\mu\nu} & = & {\bar{\sum}}_i \sum_f (2\pi)^3V \delta^4(P_f + p_{l} -
 P_i - p_\nu )
     <\!f|J^\mu|i\!><\!f|J^\nu|i\!>^*,
\end{eqnarray*}
where $V$ is the quantization volume.
In the Fermi-gas model,  $W^{\mu\nu}$ can be related
to the single-nucleon transition amplitudes,
and the relation for the case of quasi-elastic scattering
is well known~\cite{moniz}.
For a single-pion production process,
$W^{\mu\nu}$ is given as
\begin{eqnarray}
W^{\mu\nu} & = & 
 \int d\vec{p}\prime\,d\vec{k}\,d\vec{p}\, 
\theta(p_F -|\vec{p}|)\theta(|\vec{p}'|-p_F)\,
\delta^4(p + q - p' - k)
\nonumber \\
& \times&
 \frac{1}{(2\pi)^3}\frac{m_N^2}{E_N(p)E_N(p')2 E_\pi(k)}\frac{3}{4\pi p_F^3}
\nonumber \\
 & \times&
\sum_{s_N,s_{N'},i,t_N}\frac{N_{t_N}}{2}\!
<\!\!\pi^i N(p',s_{N'},t_{N'})|j^{\mu}|N(p,s_N,t_N)\!\!>\nonumber\\
& & \;\;\;\;\;\;\;\;\;\;\;\;\;\;\;\; \times
<\!\!\pi^i N(p',s_{N'},t_{N'})|j^{\nu}|N(p,s_N,t_N)\!\!>^*
\end{eqnarray}
Here 
$p$ ($p\prime$) is the four-momentum of the initial (final) nucleon,
$k$ is the four-momentum of the pion, $q=p_\nu - p_l$, 
and $p_F$ is the Fermi momentum;
$N_{t_N} = Z/N$ are the proton and neutron numbers in the target nucleus;
$s_N$ and $t_N$ are the spin and isospin of the struck nucleon,
while $i$ is the isospin index of the pion.
The matrix element of the nucleon current
for pion production, $<\!\pi^i N(p',s_{N'},t_{N'})|j^{\mu}|N(p,s_N,t_N)\!>$,
is calculated using the SL model.  
Since SL gives the
pion-production amplitude in the pion-nucleon center-of-mass frame
($\pi N$-cm frame, for short),
we transform it into the amplitude in the rest frame of the target nucleus
(LAB frame) according to
\begin{eqnarray}
W^{\mu\nu} & = & 
 \sum_{s_N,s_{N'},t_N, i} \frac{3}{4\pi p_F^3}
\int \!d\vec{p} \,\theta(p_F\! -\!|\vec{p}|)\frac{m_N}{E_N(p)}
\nonumber \\
& \times&
N_{t_N}   \int \!d\Omega_* \theta(|\vec{p}\prime|\!-\!p_F)\,
 \frac{|\vec{k}_c|m_N}{32\pi^2 W} \nonumber \\
 & \times&
\Lambda^{\mu\mu'}\!\!
<\!\pi^i N(p',s_{N'})|j^{\mu'}|N(p,s_N,t_N)\!>_{\pi N{\rm -cm}}
\nonumber \\
&\times&
\Lambda^{\nu\nu'}\!\!
<\!\pi^i N(p',s_{N'})|j^{\nu'}|N(p,s_N,t_N)\!>^*_{\pi N{\rm -cm}},
\label{wmunu}
\end{eqnarray}
where $W$ is the invariant mass of the pion and nucleon given by
$W = \sqrt{(p'+k)^2}$.
The Lorentz transformation matrix $\Lambda^{\mu\nu}$ 
transforms vectors in the $\pi N$-cm frame
to those in the LAB frame.
In the $\pi N$-cm frame,
$p' + k = (W, \vec{0}) $ ,
whereas in the LAB frame we identify
$ p' + k = p+q = (\sqrt{\vec{p}^2+m_N^2}-B + 
\omega,\vec{p}+\vec{q})$;
thus the nuclear binding correction 
is taken into account with the use of 
$p^\mu=(\sqrt{\vec{p}^2 + m_N^2}-B, \vec{p})$.
We note that the Pauli blocking factor,
$\theta(|\vec{p}\prime|-p_F)$, is dependent on the pion momentum $\vec{k}$
through $\vec{p}\prime = \vec{p}+ \vec{q} - \vec{k}$,
and hence the consideration of the Pauli blocking effect
requires the knowledge of the pion-production amplitude.
We come back to this point later.

We take into account the nuclear correlation effects 
in the initial state by using the spectral function $P(\vec{p},E)$
obtained in Ref.~\cite{benhar1}.
This is achieved by the following replacements in Eq. (\ref{wmunu})
\begin{eqnarray}
 \frac{3}{4\pi p_F^3}\int \! d\vec{p} \,\theta(p_F\! -\!|\vec{p}|)
& \rightarrow  & \int \! d\vec{p}\,dE P(\vec{p},E) \\
p^0= \sqrt{\vec{p}^2 + m_N^2}\! -\! B
& \rightarrow & m_N \!- \!E.
\end{eqnarray}
We note that $P(\vec{p},E)$ is normalized as
$\int \!d\vec{p} \,dE P(\vec{p},E) = 1$.
Although the use of the spectral function
implies that the separation of occupied and empty nucleon orbits
based on the Fermi momentum $p_F$ is no longer strictly valid,
we choose to retain the factor $\theta(|\vec{p}\prime|-p_F)$ 
in Eq.(\ref{wmunu}) to approximately take account of Pauli blocking 
for the final nucleon.

\section{Results and Discussion}

Using the formalism explained in the previous section,
we calculate neutrino-nucleus reaction cross sections
for a representative case of the $\nu$-$^{12}$C scattering.
For the incident neutrino energy we take $E_\nu=1$ GeV,
a value lying in the energy region of current importance for 
many neutrino oscillation experiments.
We first discuss the pion-production cross sections, 
which are our main results, and subsequently we consider
the combined contributions of the pion-production and quasi-elastic
processes.

\begin{figure}
\includegraphics[width=7cm]{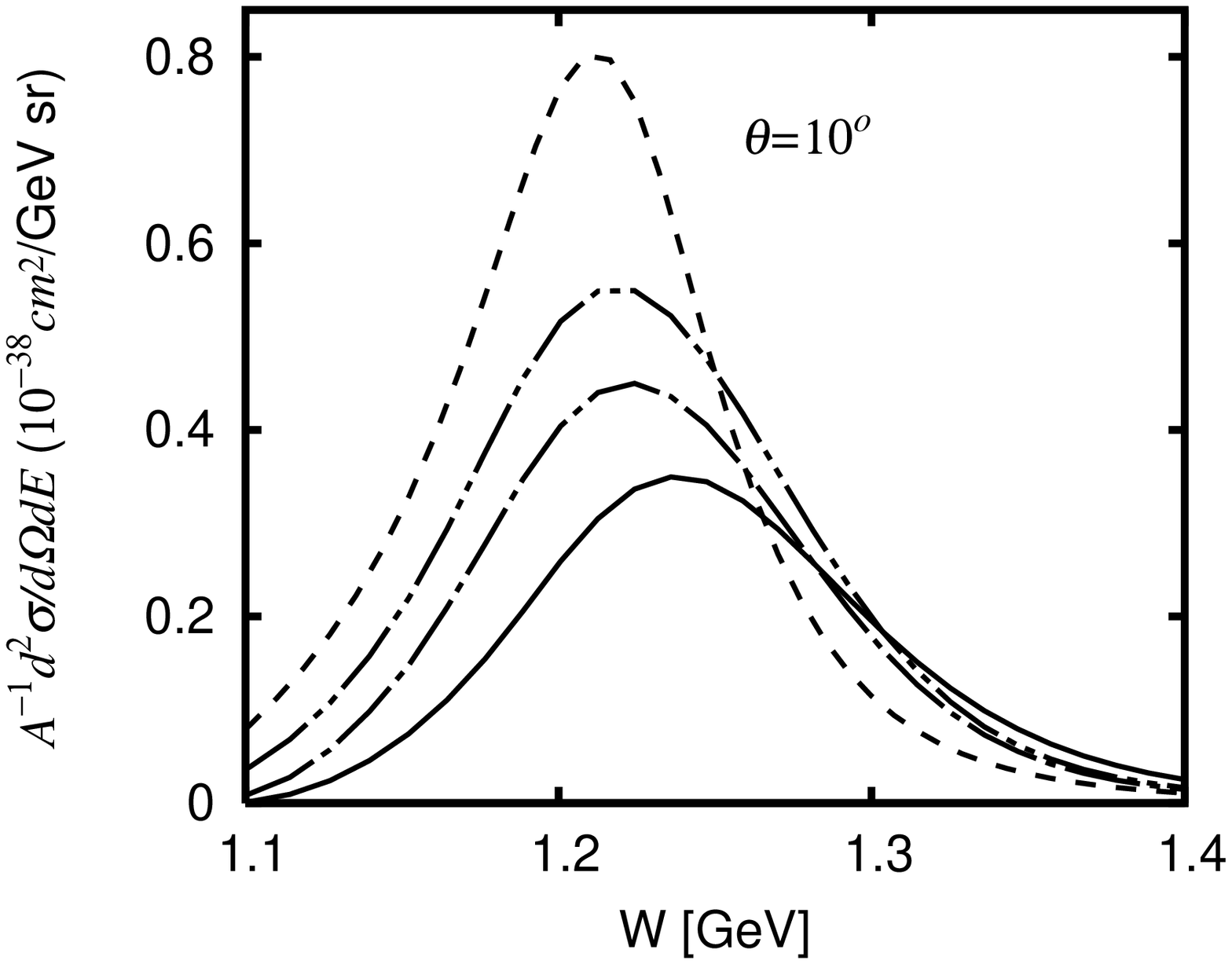}
\includegraphics[width=7cm]{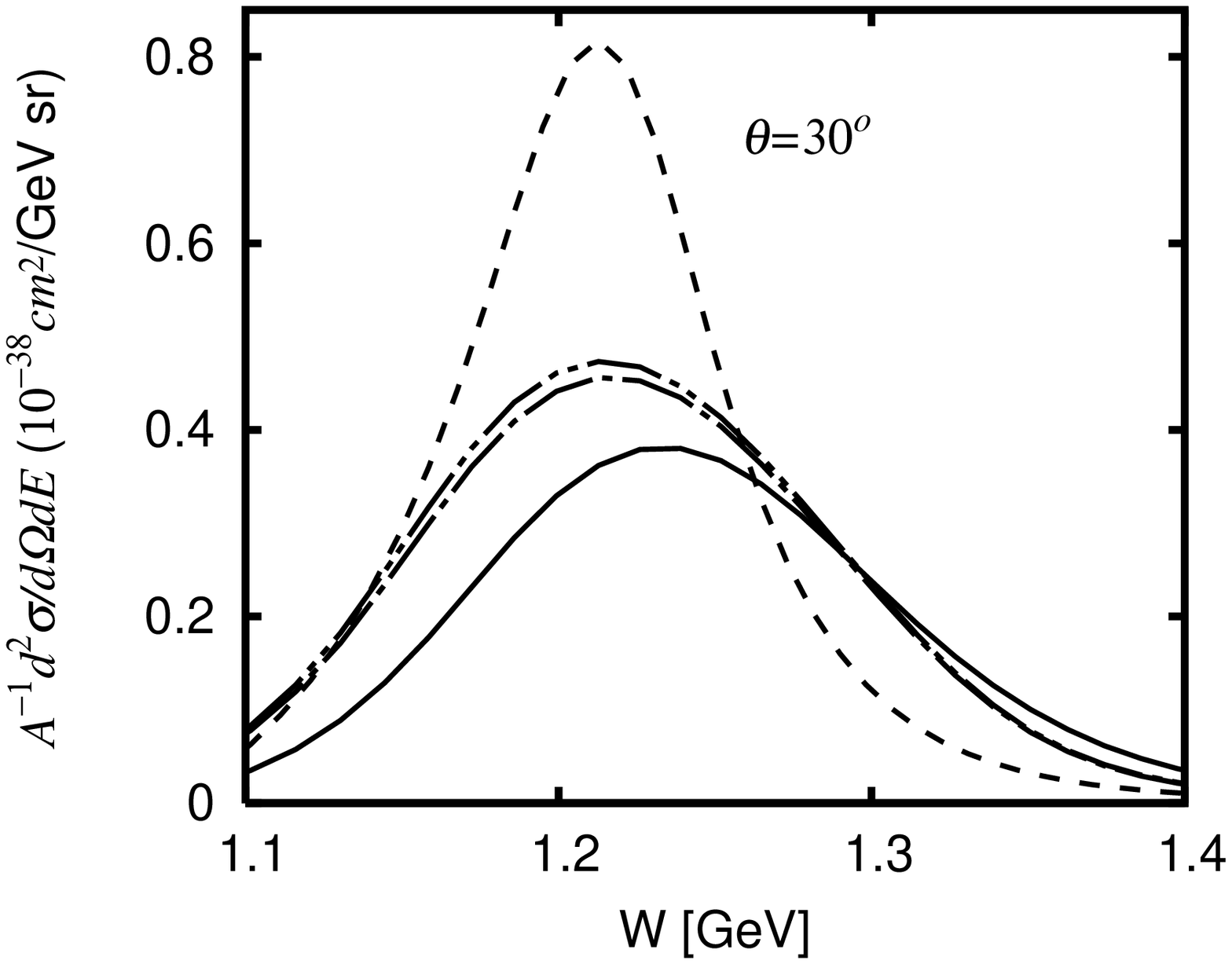}
\caption{\label{fig-cc-pi} Differential cross sections for
$ \nu_e ^{12}$C$\rightarrow e^-\pi X$
at $\theta=10^\circ$ and $30^\circ$. 
Dashed line -- free-nucleon;
dashed-double-dotted line -- Fermi-gas model;
dashed-dotted line -- Fermi-gas model with Pauli blocking; 
solid line -- full calculation.}
\end{figure}

The differential cross sections for the 
$\nu_e  ^{12}$C$\rightarrow e^- \pi  X$ reaction,
normalized with the target mass number ($A=12$),
are shown in Fig.~\ref{fig-cc-pi},
for the lepton scattering angle $\theta=10^\circ$ and $30^\circ$,  
as a function of the invariant mass 
$W=\sqrt{-Q^2+m_N^2 +2\omega m_N}$.
The dashed curve is the cross section 
obtained simply by taking the average of the 
incoherent contributions of the free protons and neutrons,
while the Fermi-gas model results are shown 
by the dash-double-dotted curves.
As expected, the inclusion of the nucleon Fermi motion 
widens the resonance width compared with the free nucleon case.
The dash-dotted curves, corresponding to the case
that includes the Pauli blocking effect for the final nucleon,
indicate that the blocking effect reduces the forward cross section
by about 20\%.
As mentioned in connection to Eq.~(\ref{wmunu}),
the inclusion of the Pauli blocking effect for the $\pi$-production process
requires the knowledge of the pion-production amplitude.
This implies that this effect cannot be evaluated
by taking (as often done in the literature) 
the incoherent sum of the free-nucleon pion-production strength
over the Fermi sea.
(The role of Pauli blocking, however,
diminishes for larger angles, 
where the momentum transfer becomes larger than the Fermi momentum.)
The solid curves show the results of our full calculation
that includes the spectral function taken from Ref.~\cite{benhar1};
it is seen that the nuclear correlation effects further
broaden the peak width and reduce the peak height by about 20\%.

\begin{figure}
\includegraphics[width=7cm]{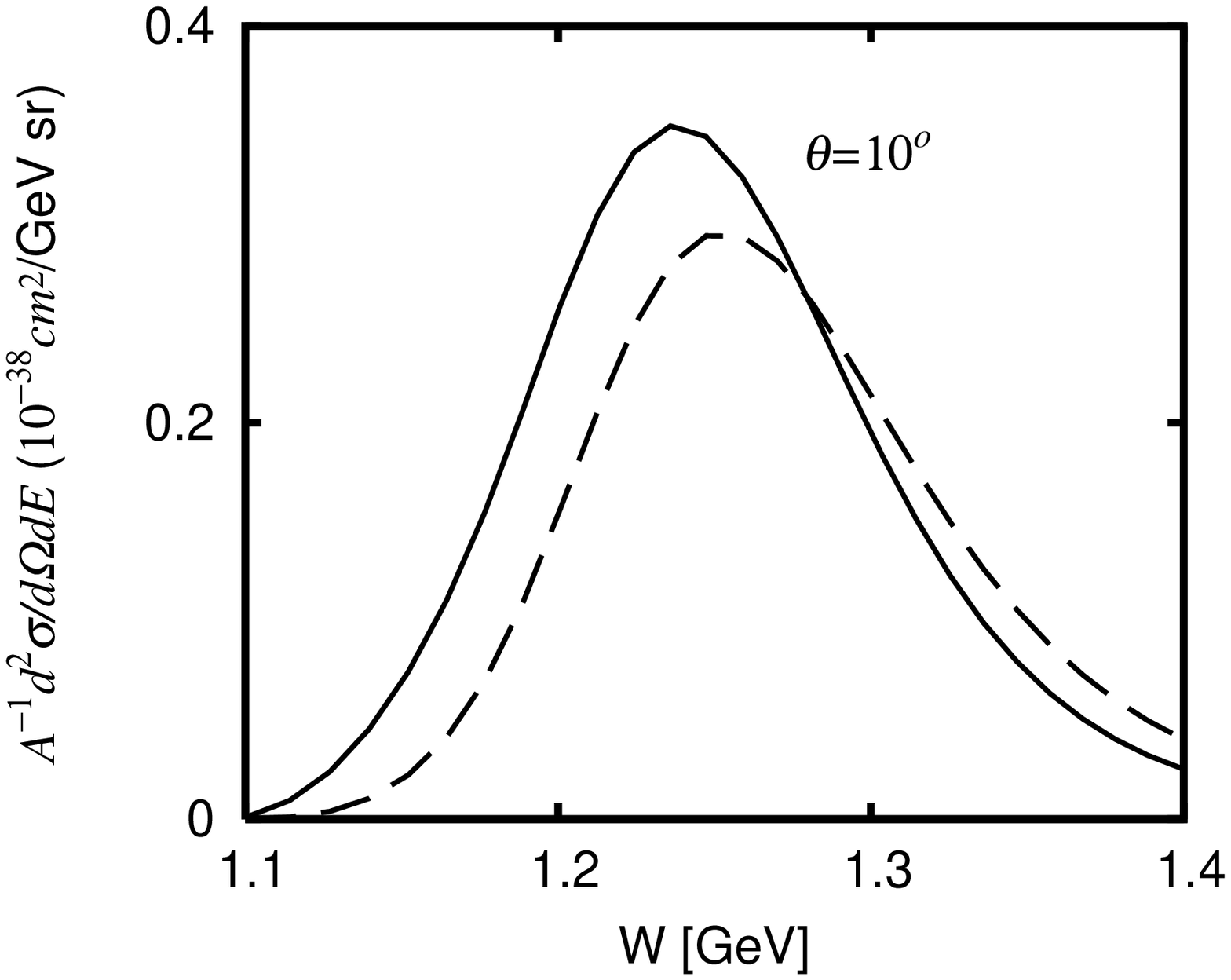}
\includegraphics[width=7cm]{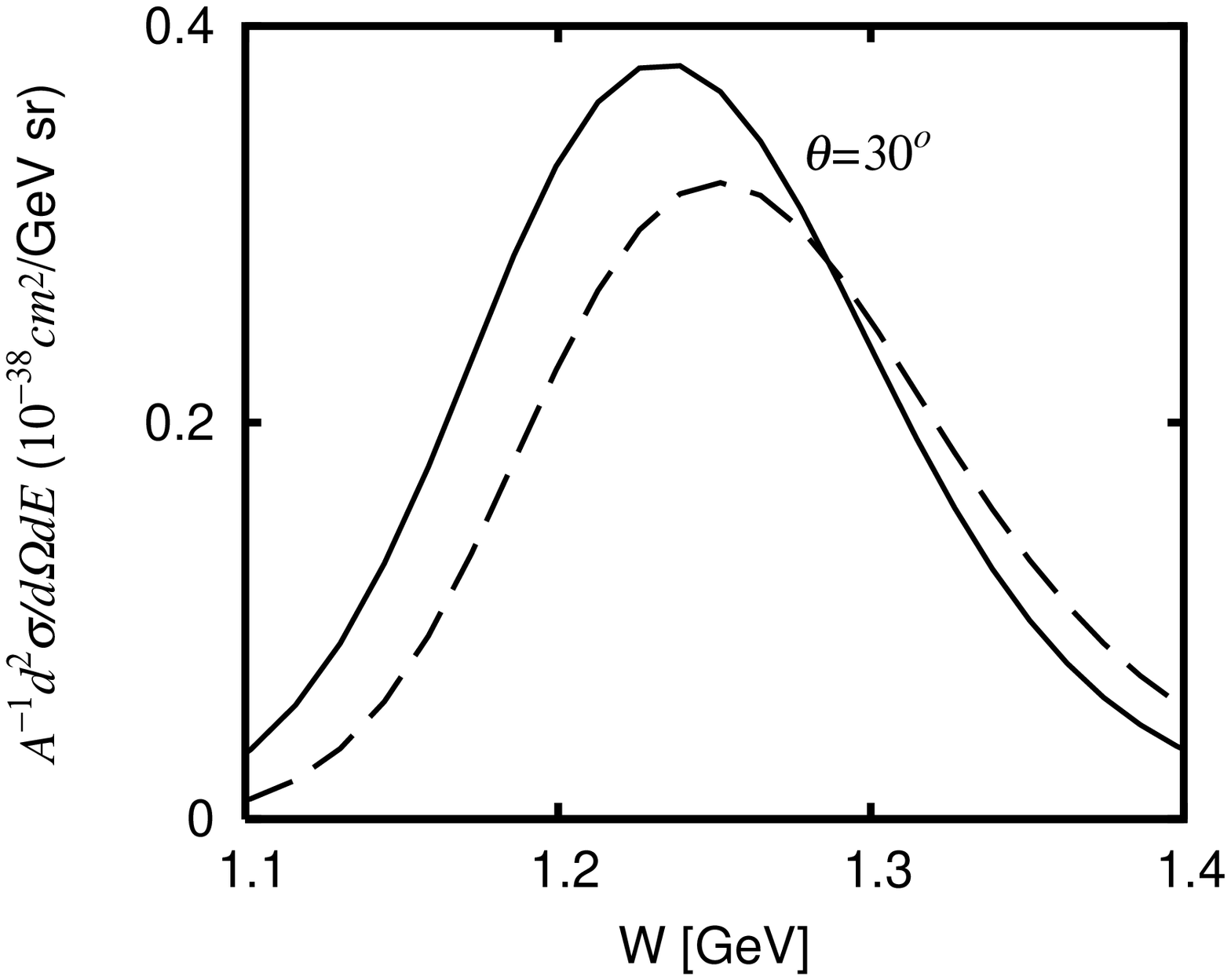}
\caption{\label{fig-cc-pi-bg} Differential cross sections for
$ \nu_e ^{12}{\rm C} \rightarrow e^- \pi X$
at $\theta=10^\circ$ and $30^\circ$.
The dashed line represents the case where 
the pion production amplitude contains
the dressed $\Delta$ contribution alone, 
while the solid line shows the results of our full calculation.}
\end{figure}

As discussed, the non-resonant mechanism 
contained in the SL-model plays a more pronounced role 
for the neutron than for the proton.
To illustrate how this feature affects the neutrino-nucleus reaction,
we give in Fig.~\ref{fig-cc-pi-bg}  
the differential cross sections 
for $ \nu_e ^{12}$C$\rightarrow e^- \pi X$
calculated with and without the non-resonant contributions;
the solid curve is the result of the full calculation,
while the dashed curve presents the case
where only the contribution of the dressed resonance amplitude
is considered.  
As expected, the resonance-only approach underestimates 
the cross sections by about 20\% even in the resonance region.
The full calculation is found to give more strength for lower values 
of $W$ than the resonance-only case.

\begin{figure}
\includegraphics[width=7cm]{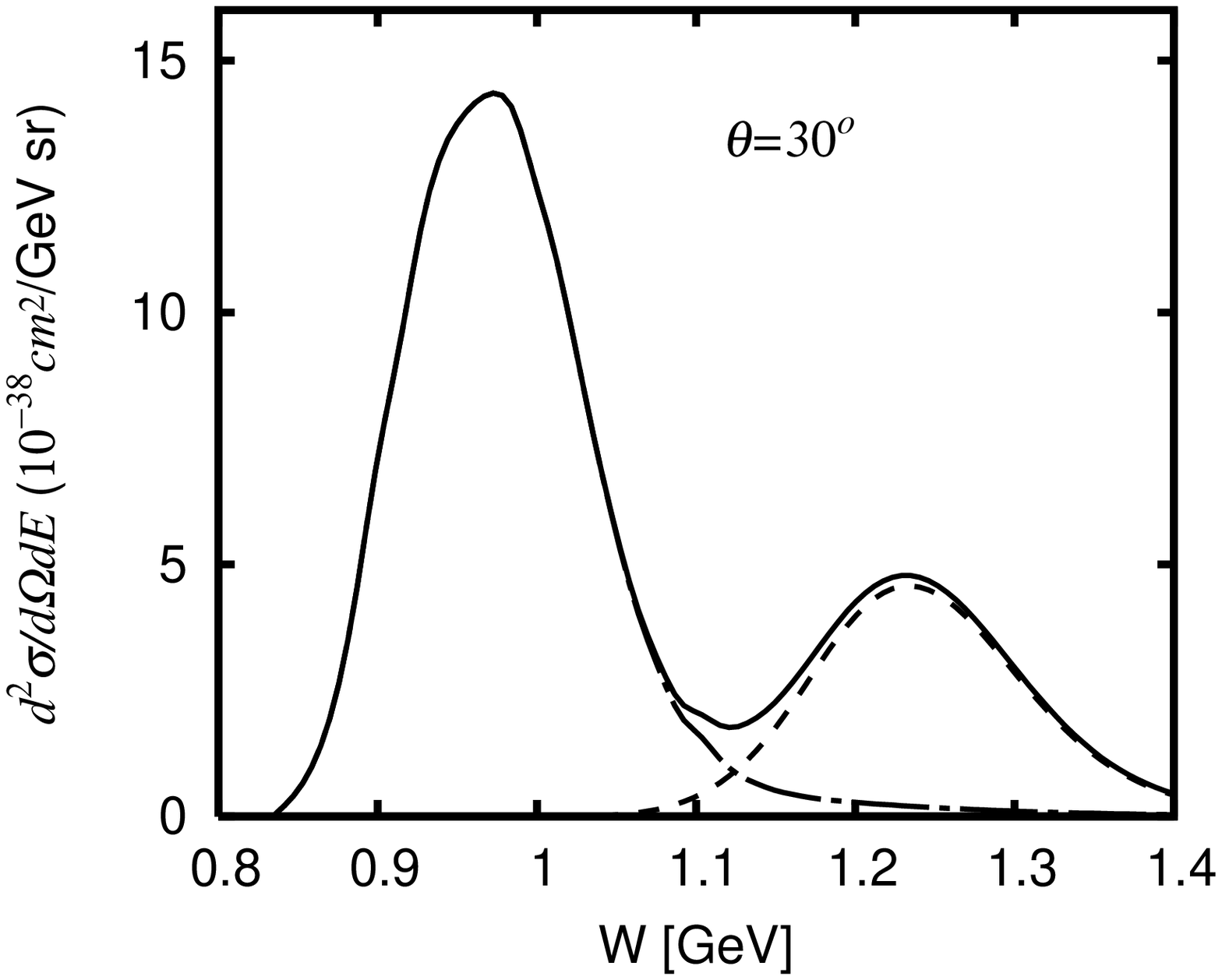}
\includegraphics[width=7cm]{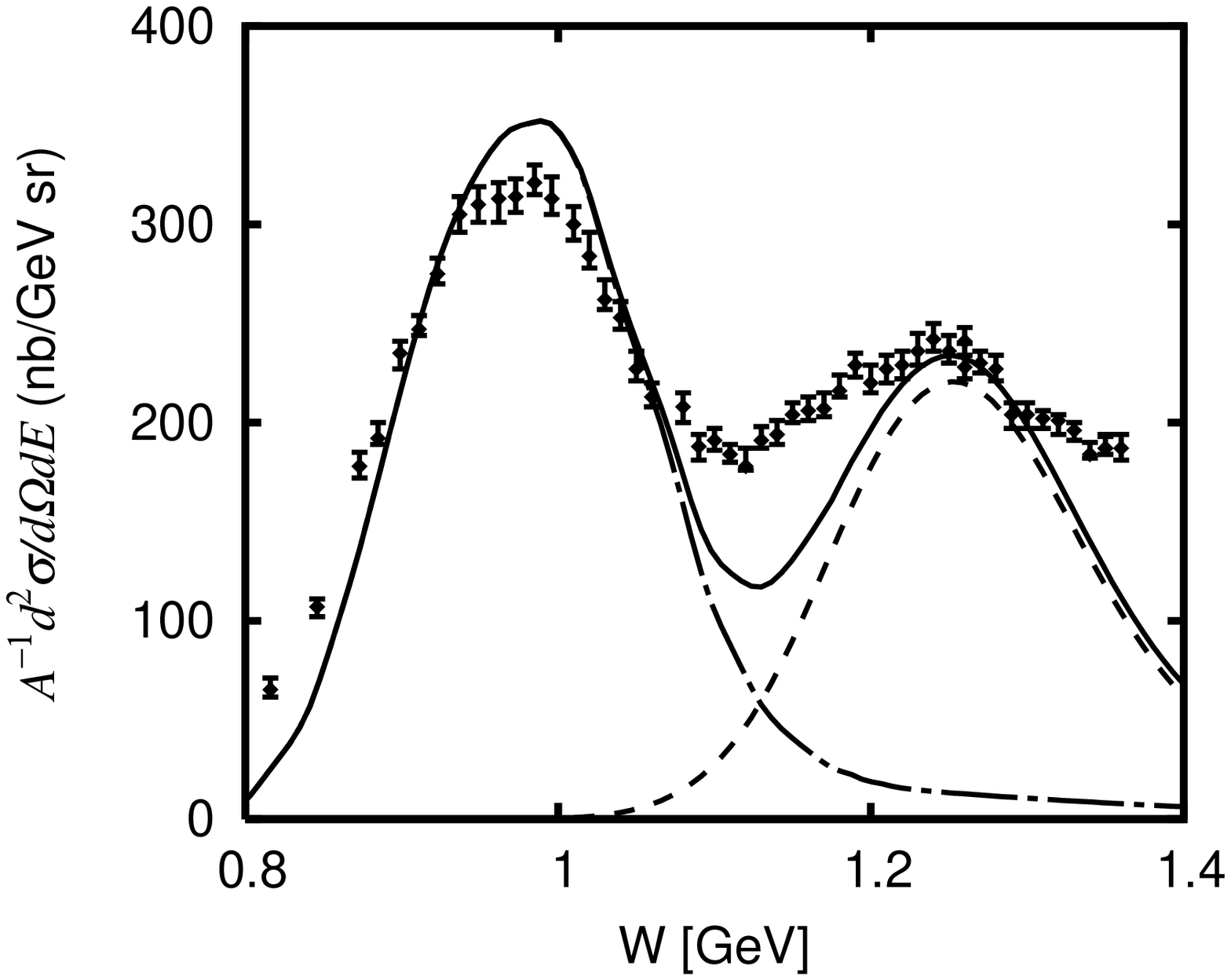}
\caption{\label{fig-cc-em} 
Left panel -- Differential cross section for 
the $\nu_e ^{12}$C$\rightarrow e^- X$ reaction
at $E_\nu=1$ GeV and $\theta=30^\circ$.
Right panel -- Differential cross section for the 
${e^-}\!+\! ^{12}{\rm C}\rightarrow e^- X$ reaction 
at $E_e=1.1$ GeV with $\theta_e=37.5^\circ$.
The experimental data points are from Ref.~\cite{sealock}.}
\end{figure}

\begin{figure}
\includegraphics[width=7cm]{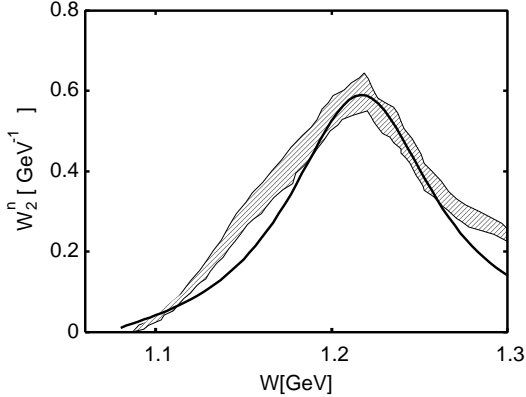}
\caption{\label{fig-w2n} The neutron structure function $W_2^n$ at
$E_e=2.445GeV$ and  $\theta_e=20^o$. The solid line represents the results of 
the SL-model and the shaded area represents the $W_2^n$
from the analysis of  Ref.~\cite{benhar06}.
}
\end{figure}

To the contribution of the pion-production process we now add
the contribution of the quasi-free nucleon knockout process.
The latter is calculated using again
the modified Fermi-gas model that 
incorporates the spectral function~\cite{Benetal05,benhar1}.
At the incident energy under consideration,
the sum of these two contributions is expected to
give the bulk of the inclusive reaction cross section.
The differential cross section for
$\nu_e ^{12}$C$\rightarrow e^- X$
at $\theta=30^\circ$
is shown in the left panel in Fig.~\ref{fig-cc-em}.
%
The bump at the lower energy is due to quasi-free nucleon knockout,
while the higher energy bump is due to $\Delta$-resonance excitation.
To examine the validity of our present approach, 
we apply the same calculational framework 
to the $e^- \!+\!^{12}{\rm C} \rightarrow e^- X$ reaction
(with the weak current replaced by the electromagnetic current),
and compare the results with the experimental data.
Fig.~\ref{fig-cc-em} shows this comparison.
It is seen that the general trend of the data is reproduced
reasonably well; in particular, the magnitude of the cross section 
in the $\Delta$-resonance region is well reproduced.
We remark that a calculation by Benhar {\it et al.}~\cite{Benetal05}(BFNSS)
underestimates the height of the $\Delta$ peak. 
According to Ref.~\cite{benhar06},  this is perhaps
mainly due to the fact that, around the $\Delta$ region,
the neutron structure functions of Ref.~\cite{bodek81}
used in BFNSS are significantly weaker 
than those extracted from an analysis of inclusive electron scattering 
on the deuteron.
It is noteworthy that  the SL model, 
in addition to providing a satisfactory description 
of the proton structure functions~\cite{sl4},
gives neutron structure functions that 
are in good agreement with those deduced in Ref.~\cite{benhar06};
see Fig.~\ref{fig-w2n}.
This feature explains the difference in the $\Delta$ peak height
between our results and those of BFNSS.
%

We note, however, that our calculation gives a dip structure
that is somewhat too deep, a feature that seems to indicate 
that we need to go beyond  `impulse' approximation
and/or employ more elaborate treatments of nuclear correlation effects;
see also Ref.~\cite{gil97}.
It is also to be noted that in the higher $W$ region
our model, which only includes the $\Delta$-resonance, 
is likely to underestimate the transition strength.

\begin{figure}
\includegraphics[width=7cm]{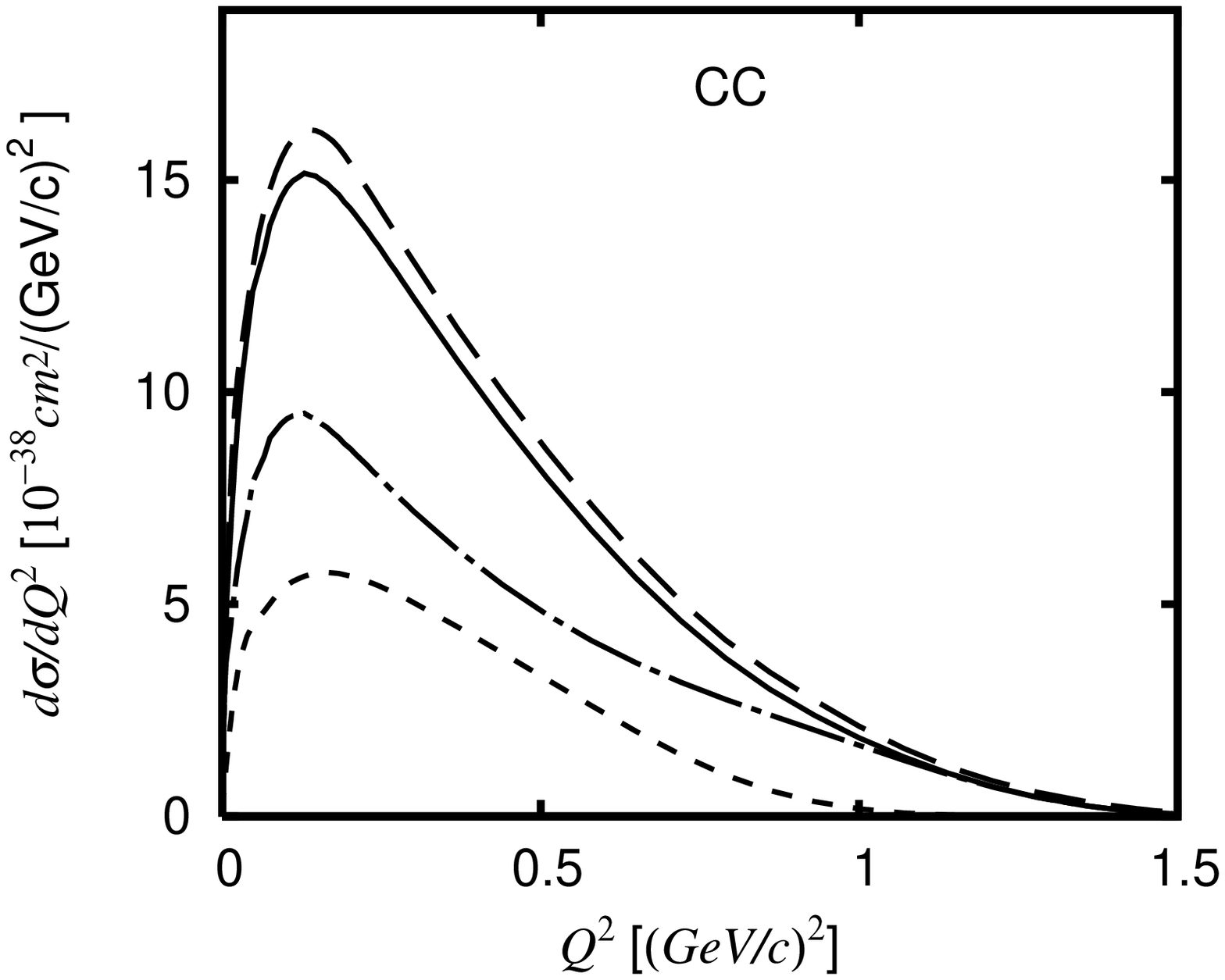}
\includegraphics[width=7cm]{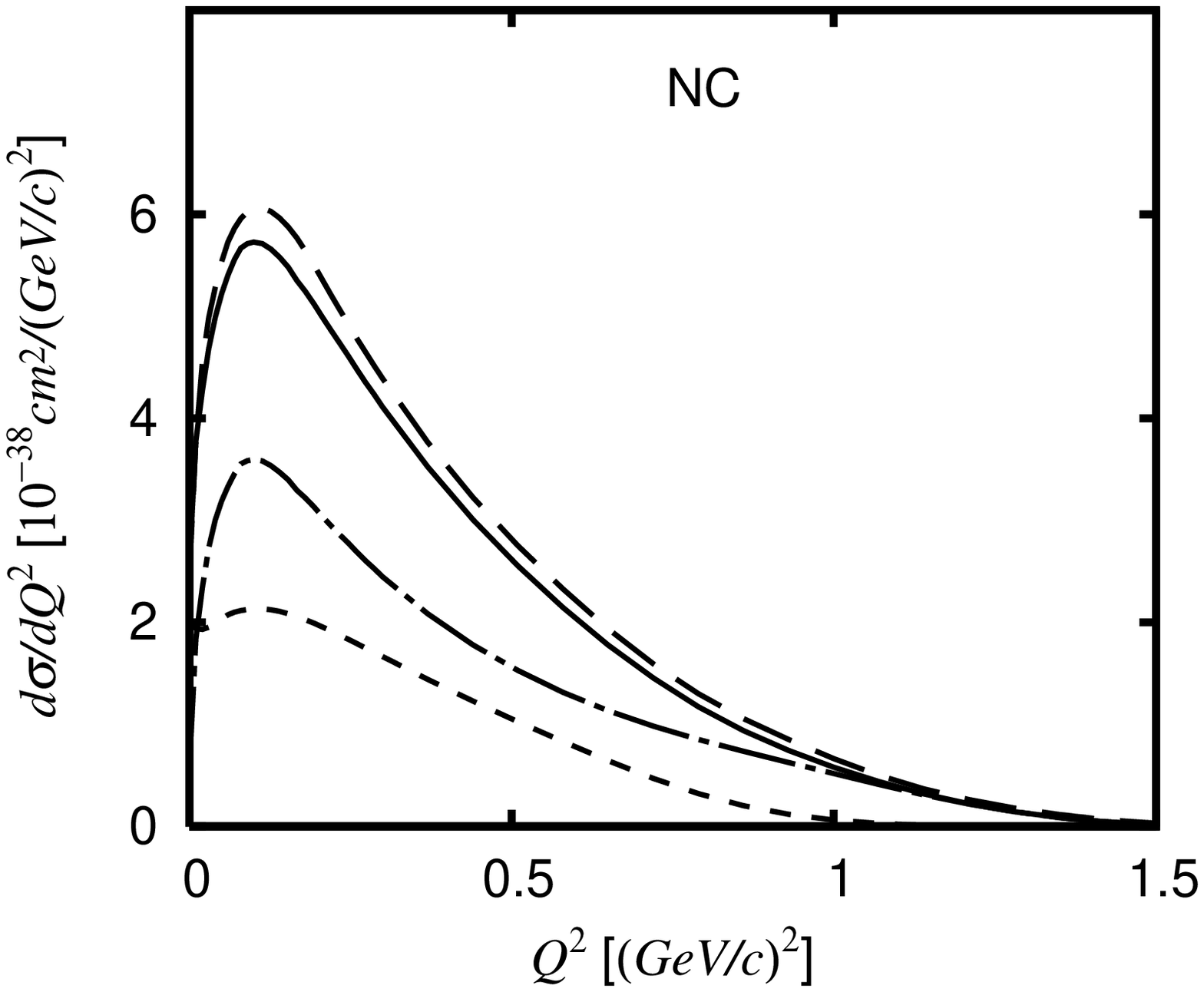}
\caption{\label{fig-dq} The differential cross section $d\sigma/dQ^2$ for 
$ \nu_\mu ^{12}$C$\rightarrow \mu^- X$ and
$ \nu ^{12}$C$\rightarrow \nu X$ at $E_\nu=$1~GeV.
The sum of the quasi-free and pion-production 
contributions is shown by the solid line (full calculation),
and by the long-dashed line (Fermi-gas model).
The individual contribution of the quasi-free process 
is shown by the dash-dotted line,
and that of the pion-production process by the short-dashed line.
}
\end{figure}

\begin{figure}
\includegraphics[width=7cm]{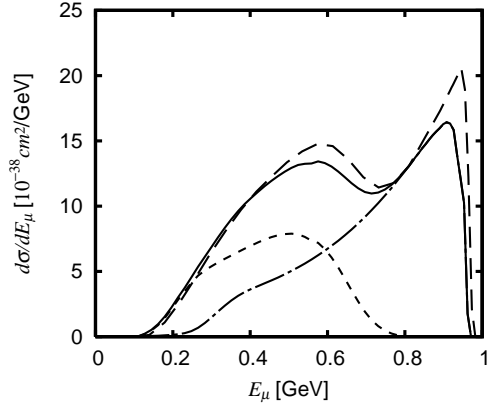}
\caption{\label{fig-de} Muon energy distribution $d\sigma/dE_{\mu}$ for 
the $\nu_\mu ^{12}$C$\rightarrow \mu^- X$ reaction.
The contributions of the quasi-free and pion-production reactions are
shown in the short-dashed and dash-dotted lines, respectively,
while their sum is given by the solid line.
The long-dashed line represents the results obtained 
in the Fermi-gas model.}
\end{figure}

For some purposes it seems useful to 
present our results in the form of $Q^2$-distribution 
[$Q^2=-(p_\nu - p_{l})^2$] or $E_{l}$-distribution. 
We get $d\sigma/dQ^2$ and $d\sigma/dE_l$ 
using the formulas,
\begin{eqnarray}
\frac{d\sigma}{d Q^2} & = & \int \!\!d E_{l} \frac{\pi}{p_\nu p_{l}}
\frac{d\sigma}{d\Omega_l dE_{l}}\;\;,\;\;\;\;\;\;\;\;
\frac{d\sigma}{d E_{l}} =  \int \!\! d\Omega_l
\frac{d\sigma}{d\Omega_l dE_{l}}.
\end{eqnarray}
Fig.~\ref{fig-dq} gives the $Q^2$ and $E_\mu$ spectra for the
$\nu_\mu ^{12}$C reaction at $E_\nu=1$ GeV;
the left (right) panel corresponds to the CC (NC) reaction.
The total contribution (solid line) consists of 
the  quasi-free contribution (dash-dotted line) and
the pion production contribution (short-dashed line).
We note that the pion production contribution 
is reduced by the introduction of the spectral function.
Finally, the muon energy distribution $d\sigma/dE_{\mu}$ in the
$ \nu_\mu ^{12}$C$\rightarrow \mu^- X$ reaction is shown in 
Fig.~\ref{fig-de}.

\section{Summary}
We have studied the neutrino-nucleus and electron-nucleus reactions 
in the $\Delta$-resonance region with the use of
the Sato-Lee (SL) model, which allows us to
treat the resonant and non-resonant contributions in a unified manner
in deriving the amplitudes for pion electroweak production on a nucleon.
The validity of SL has been extensively tested 
for electromagnetic observables involving single-nucleon targets,
and we can use the available electron-nucleus scattering data
to assess the reliability of the application of SL 
to neutrino-nucleus reactions.  
As for the nuclear correlation effects,
we have considered here only those effects
which can be considered to be subsumed in the spectral function.   
Despite this rather limited treatment 
of the nuclear effects,
our calculation based on SL
gives reasonably good descriptions of
the relevant electron scattering data; 
the peak structure in the cross section
in the resonance region is well reproduced by our calculation.
It is reasonable to expect that our calculation 
of the neutrino-nucleus reaction cross sections based on
the SL model enjoys the same level of success.
It seems worthwhile to further develop SL studies
of neutrino- and electron-nucleus reactions
by elaborating the treatment of the nuclear effects
(including medium effects on the $\Delta$-resonance itself).

\section*{Acknowledgments}
The authors are grateful to M. Sakuda for useful discussions.
This work is supported 
by the U.S. National Science Foundation, Grant No. PHY-0457014,
by the Japan Society for the Promotion of Science
Grant-in-Aid for Scientific Research(c) 15540275,
and by the U.S. Department of Energy,
Nuclear Physics Division Contract No. DE-AC02-06CH11357.

\end{document}